\documentclass[11pt]{article}
\usepackage[final]{acl}

\usepackage{caption}
\usepackage{times}
\usepackage{latexsym}
\usepackage[T1]{fontenc}
\usepackage[utf8]{inputenc}
\usepackage{microtype}
\usepackage{inconsolata}
\PassOptionsToPackage{hyphens}{url}

\usepackage{amsmath}
\usepackage{stfloats}
\usepackage{threeparttable}
\usepackage{booktabs}
\usepackage{tabularx}
\usepackage{array}
\usepackage{float}
\usepackage{graphicx}
\usepackage{bbm}
\usepackage{dsfont}
\usepackage{bm}

\usepackage{amssymb}
\usepackage{chngcntr}
\usepackage{algorithm}
\usepackage{algpseudocode}
\usepackage{subcaption}
\usepackage{xcolor}
\usepackage{multirow}
\usepackage{enumitem}
\usepackage{listings}
\usepackage{etoolbox}
\usepackage{soul}
\usepackage{xspace}
\usepackage{cuted}

\sethlcolor{yellow}

\microtypesetup{protrusion=true, expansion=true}
\newcommand{\sysname}{Proposed Model\xspace}

\soulregister{\sysname}{0}
\soulregister{\_}{0}
\soulregister{\&}{0}
\soulregister{\%}{0}
\soulregister{\sim}{0}
\soulregister{\textbf}{1}
\soulregister{\textit}{1}
\soulregister{\textsc}{1}

\newcounter{stepnum}
\counterwithin{stepnum}{algorithm}

\DeclareCaptionLabelFormat{plain}{#1 #2}

\begin{document}

\title{Semantic Retrieval for Product Search in E-Commerce\vspace{10pt}}
\author{
  \textbf{Nikhil Kothari} \And
  \textbf{Saksham Samdani} \And
  \textbf{Ritam Mallick} \AND
  \textbf{Praveen Gupta} \And
  \textbf{Ankit Vijay} \And
  \textbf{Surender Kumar} \AND
  Flipkart, India \\
  \texttt{\{nikhil.kothari, saksham.samdani, ritam.mallick,} \\
  \texttt{gupta.praveen, ankit.vijay, surender.k\}@flipkart.com}
}
\maketitle

\setlength{\stripsep}{0pt}
\begin{strip}
\noindent
\begin{minipage}[t]{\columnwidth}
\begin{abstract}

Semantic retrieval in e-commerce must handle short, noisy, and colloquial queries over large product catalogs with fine-grained attribute distinctions. We present a Siamese LLM dual-encoder trained through a two-stage pipeline: contrastive learning with a \textit{false-negative margin mask} to prevent penalization of near-duplicate products, followed by \textit{Relative Odds Alignment for Retrieval} (ROAR), a preference optimization objective that extends Bradley--Terry to variable-sized graded relevance groups via consecutive odds-ratio margins. The training corpus mirrors this progression — substitute query--product pairs provide coarse semantic supervision in Stage~1 and graded relevance annotations drive fine-grained ranking in Stage~2. The resulting system accurately retrieves exact matches while correctly ordering substitutes and complementary products, with gains confirmed across query-frequency strata and business verticals, and statistical significance validated through live A/B deployment at scale.

\end{abstract}
\end{minipage}%
\hfill
\begin{minipage}[t]{\columnwidth}
% Align figure top with abstract text start.
\vspace{\dimexpr\topsep+\parskip\relax}
\centering

\begin{minipage}[c][0.30\textheight][c]{1.0\columnwidth}
  \centering
  \includegraphics[
    width=\linewidth,
    height=0.30\textheight,
    keepaspectratio,
    trim={22pt 10pt 20pt 10pt},
    clip
  ]{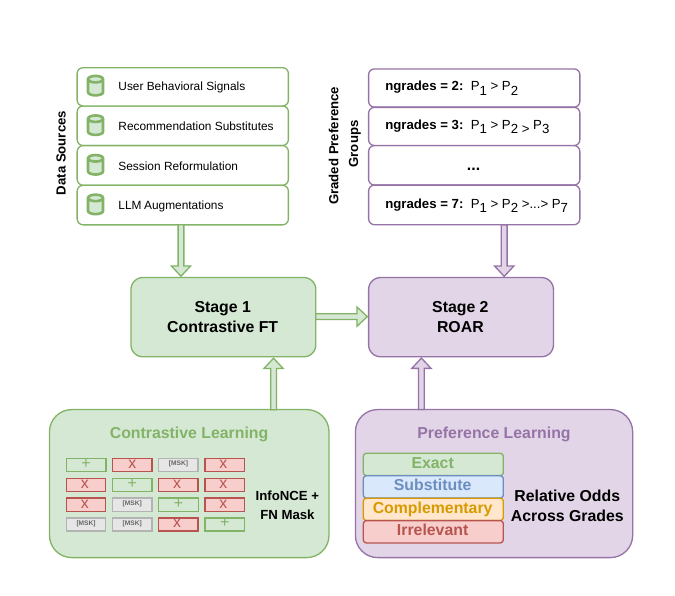}
\end{minipage}

\captionof{figure}{\textmd{Two-stage training pipeline. \textit{Stage~1}: contrastive fine-tuning of Qwen3-Embedding-4B. \textit{Stage~2}: Relative Odds Alignment for Retrieval over variable-size graded groups.}}
\label{fig:pipeline}

\end{minipage}
\end{strip}

\vspace*{-1.5ex}
\section{Introduction}

Search is the primary tool for customers to discover products in e-Commerce, so ensuring high relevance of search is critical to user satisfaction and trust. If search results do not match the intent of the search -- for example, showing wrong brands or completely irrelevant item types -- it damages the shopping experience and business results~\citep{li2021learning}. The system must exhibit world knowledge, understand functional intent, honor category boundaries in substitutes, and handle brand--category ambiguity---all while serving diverse product verticals from electronics to fashion. Recent advances in \textit{deep learning for semantic search} offer new opportunities to improve retrieval quality beyond simple lexical matching. In particular, LLM-based language models have shown strong capability to capture query--item semantics. For example, previous work used a cross-encoder model to compute query product relevance with high precision~\citep{nogueira2019passage}. However, cross encoders are computationally expensive for the first-stage retrieval of millions of products, as they must jointly encode each query with each candidate~\citep{karpukhin2020dense}. To ensure efficiency, many production systems instead use dual encoder (two tower) models that encode the query and the product separately in a shared embedding space~\citep{karpukhin2020dense}. This allows pre-computing product embeddings and fast retrieval via nearest-neighbor search, at the cost of some loss in precision compared to cross-encoders~\citep{karpukhin2020dense}.

\textbf{Our objective is to train an embedding-based retrieval model that captures query--product relevance with high fidelity}, bridging the gap between cross-encoder accuracy and dual-encoder efficiency. To this end, we combine two training paradigms: (1) \textit{Contrastive learning} to learn a unified embedding space for queries and products, and (2) \textit{Preference alignment fine-tuning} to directly optimize ranking order according to relevance feedback. Stage-1 applies a contrastive loss with random in-batch negatives, Stage-2 then applies a Preference Alignment loss to fine-tune the model on multi-graded preferences (Exact Match > Substitutes > Complementary > Irrelevant Items).

\vspace*{3pt}
\textbf{Contributions.} The main contributions of this work include:
\begin{itemize}

\item \textbf{Search Relevance:}

We present a two-stage training framework for building an LLM-based relevance system at production scale, and validate its effectiveness through extensive offline evaluation and online A/B testing. The second stage introduces \textit{Relative Odds Alignment for Retrieval} (ROAR), a novel preference-optimization objective for retrieval that supports a variable number of graded relevance labels per query and directly optimizes the relative ordering of products within each query group. To the best of our knowledge, this is the first application of odds-ratio-based preference optimization to graded-relevance retrieval.

\item \textbf{Dataset:}
We demonstrate the benefit of a large-scale, multi-source, multilingual training corpus, constructed from real search logs, human-annotation, behavioral interaction signals, recommendation-derived substitutes, session-based query reformulations, and LLM-driven augmentations.
\end{itemize}

\section{Related Work}

\subsection{Dense Retrieval}

Dense retrieval encodes queries and documents into a shared embedding space and retrieves candidates via maximum inner-product or cosine similarity search. DPR~\citep{karpukhin2020dense} established the dual-encoder paradigm for open-domain QA using BERT-based encoders with in-batch negatives. Subsequent work scaled this approach: Sentence-BERT~\citep{reimers2019sentence} introduced Siamese training for general-purpose sentence embeddings, while ColBERT~\citep{khattab2020colbert} proposed token-level late interaction for higher precision at moderate cost. GTR~\citep{ni2022large} demonstrated that scaling T5 encoders to multi-billion parameters improves generalization across diverse tasks.

More recently, decoder-only LLMs have been repurposed as embedding models. E5-Mistral~\citep{wang2024improving} and NV-Embed~\citep{lee2024nv} showed that instruction-tuned LLMs produce competitive embeddings via last-token pooling. GritLM~\citep{muennighoff2024generative} unified generation and embedding in a single model. The Qwen embedding family~\citep{qwen2025embedding} extends this line by training decoder-only models specifically for embedding tasks with multi-stage contrastive and preference-based objectives. Our work builds on this foundation, employing Qwen3-Embedding as the Siamese LLM backbone with task-specific training for e-commerce retrieval.

\subsection{E-Commerce Product Search}

Product search introduces domain-specific challenges: short and noisy queries, enormous catalogs with fine-grained distinctions, diverse intents, and vocabulary mismatch between user language and seller-authored product text~\citep{li2021learning}. Embedding-Based Retrieval (EBR) at Facebook Marketplace~\citep{huang2020embedding} demonstrated production-scale two-tower deployment with unified text and image embeddings. Nigam et al.~\citep{nigam2019semantic} at Amazon showed the importance of query--product type understanding for e-commerce. MOBIUS~\citep{fan2019mobius} at Baidu proposed a teacher-student framework for retrieval at billion-scale.

\subsection{Datasets}

For e-commerce specifically, the Amazon Shopping Queries Dataset~\citep{reddy2022shopping} provides 130K+ queries with ESCI (Exact, Substitute, Complement, Irrelevant) relevance labels across multiple locales---the closest public analogue to the graded relevance annotations used in our work. The Wands dataset~\citep{chen2022wands} offers product search relevance judgments at smaller scale.

Training data strategies are equally critical. Click-through data provides weak supervision at scale~\citep{yao2021learning}, though it requires debiasing for position and presentation effects~\citep{joachims2017unbiased}. Hard negative mining from model-based retrieval~\citep{xiong2020approximate, qu2021rocketqa} improves contrastive training quality. More recently, LLM-based synthetic data generation has emerged as a powerful augmentation strategy: InPars~\citep{bonifacio2022inpars} uses LLMs to generate synthetic queries for passages, while Promptagator~\citep{dai2023promptagator} creates few-shot task-specific training data. 

\section{Methodology}

Figure~\ref{fig:pipeline} provides an overview of the two-stage training pipeline and online deployment. The remainder of this section details each component.

\subsection{Model Architecture}
\label{sec:architecture}
We adopt a Siamese dual-encoder where query and product towers share a single Qwen3-Embedding-4B~\citep{qwen2025embedding} backbone — a decoder-only transformer used as an encoder, consistent with recent embedding model designs.

\paragraph{\normalfont\textit{Pooling.}} We use \textit{last-token (<eos>) pooling} with left-padded inputs. For each input sequence, the hidden state at the final token position is extracted as the sequence embedding. 

\paragraph{\normalfont\textit{Parameter-Efficient Fine-Tuning.}} We fine-tune with LoRA~\citep{hu2022lora} with rank $r = 32$ and scaling factor $\alpha = 64$, applied to all attention and MLP projection layers with a dropout of 0.1.

\paragraph{\normalfont\textit{Matryoshka Representation Learning.}} We adopt MRL~\citep{kusupati2022matryoshka} to jointly train nested subspace embeddings, enabling deployment at 256 embedding dimension without retraining.

\subsection{Contrastive Learning}
\label{sec:contrastive}

The first training stage optimizes the model with the improved InfoNCE-style contrastive objective used in Qwen3 Embedding~\citep{qwen2025embedding}. This objective augments standard InfoNCE by aggregating multiple negative pools, including hard negatives, in-batch queries, and in-batch documents, while using a masking term to suppress potential false negatives.

% \paragraph{False-Negative Margin Masking.}
\paragraph{\normalfont\textit{False-Negative Margin Masking.}}
A batch may contain multiple products genuinely relevant to a given query (e.g., size or color variants). Treating these as negatives introduces noisy gradients. We address this with a margin-based masking mechanism:
\begin{equation}
m_{ij} = \mathds{1}\bigl[\cos(q_i, d_j) \leq \cos(q_i, d_i) + \delta\bigr],
\label{eq:margin_mask}
\end{equation}
where $\delta = 0.1$. Any in-batch product whose similarity to $q_i$ exceeds that of the labeled positive by more than $\delta$ is excluded from the denominator as a potential false negative.

\subsection{Relative Odds Alignment for Retrieval}
\label{sec:alignment}

We further refine the model with an objective that directly optimizes graded relevance ordering. To support a variable number of relevance grades across queries, we introduce ROAR, a retrieval alignment objective that extends the Bradley--Terry pairwise comparison framework~\citep{bradley1952rank} to variable-sized graded relevance groups via consecutive odds-ratio preference margins following~\citet{hong2024orpo}.

\paragraph{Group Preference Construction.}

Each training example consists of a query $q_i$ and an ordered group of $g_i$ products ranked by relevance: $p_{i,1} \succ p_{i,2} \succ \cdots \succ p_{i,g_i}$, our formulation supports variable group sizes ($g_i \in [2, 7]$) drawn from graded relevance annotations. The gradation hierarchy reflects real retrieval requirements. For example, a 4-grade hierarchy is:
\begin{center}
\small

    \textit{Perfect Match} $\succ$ \textit{Substitute} $\succ$ \textit{Complementary} $\succ$ \textit{Irrelevant},

\end{center}

This enables the model to learn fine-grained relevance distinctions: for the query \emph{``Nandini Milk,''} the ideal retrieval ordering is exact Nandini milk products $\to$ other brand milks $\to$ other Nandini brand products $\to$ other milk-based products---a gradation that binary preference cannot capture.

\paragraph{Alignment Loss via Relative Odds.}

We formulate the alignment objective using odds ratios computed from query--product contrastive probabilities. Given a batch of $N$ queries and $M=\sum_i g_i$ candidate products, we first compute the query--product score matrix $S \in \mathbb{R}^{N \times M}$. The probability of selecting product $p_j$ for query $q_i$ is obtained using a row-wise softmax with alignment temperature $\tau_a$:
\begin{equation}
P_\theta(p_j \mid q_i)
=
\operatorname{softmax}(S_i / \tau_a)_j .
\end{equation}

For each query $q_i$, let $s_i = \sum_{r<i} g_r$ denote the starting index of its associated product group. Within this group, products are ordered by relevance, and we compare each consecutive pair. For compactness, we denote:

\begin{equation}
\begin{aligned}
p_{i,k}^{+} &= p_{s_i+k}, \quad p_{i,k}^{-} = p_{s_i+k+1},\\
k &= 0,\ldots,g_i-2.
\end{aligned}
\end{equation}

The odds of selecting a product $p_j$ for query $q_i$ is defined as:
\begin{equation}
\operatorname{odds}_\theta(p_j \mid q_i)
=
\frac{
P_\theta(p_j \mid q_i)
}{
1 - P_\theta(p_j \mid q_i)
}.
\end{equation}

The log odds-ratio preference margin between a higher-relevance product $p_{i,k}^{+}$ and the next lower-relevance product $p_{i,k}^{-}$ is then:
\begin{equation}
\Delta^{\text{OR}}_{i,k}
=
\log
\frac{
\operatorname{odds}_\theta(p_{i,k}^{+} \mid q_i)
}{
\operatorname{odds}_\theta(p_{i,k}^{-} \mid q_i)
}.
\end{equation}

Finally, the alignment loss maximizes the preference for higher-relevance products over lower-relevance consecutive products:
\begin{equation}
\mathcal{L}_{\text{align}}
=
-\frac{1}{C}
\sum_i
\sum_{k=0}^{g_i-2}
\log \sigma
\left(
\Delta^{\text{OR}}_{i,k}
\right),
\end{equation}
where $C=\sum_i (g_i-1)$ is the total number of valid consecutive product pairs.

\paragraph{\normalfont\textit{Combined Objective.}}

The alignment stage combines the alignment loss with the contrastive loss of section \ref{sec:contrastive}. For the contrastive term, the first product in each group is treated as the positive. Given group sizes $[g_0, g_1, \ldots, g_{B-1}]$, the positive index for row $i$ is the starting offset of its group:
\begin{equation}
p_0 = 0, \qquad p_i = p_{i-1} + g_{i-1}, \quad i=1,\ldots,B-1.
\end{equation}
The final objective is
\begin{equation}
\mathcal{L}_{\text{total}} = \mathcal{L}_{\text{contrast}} + \beta \cdot \mathcal{L}_{\text{align}},
\end{equation}
where $\beta$ controls the alignment loss contribution.

\section{Datasets}

Our semantic search engine is trained and evaluated on a large-scale proprietary dataset collected from our platform’s search logs. This dataset comprises millions of real user queries, associated clicked products, and various metadata. We leverage both raw logged data and numerous augmentation techniques to enrich the dataset, ensuring broad coverage of query formulations and robust query-product relevance signals. In the following section, we describe the dataset construction and the augmentation methods employed.

\subsection{Human-Annotated Data}

% \paragraph{Graded Relevance Annotations.} 
We curate a human-annotated dataset of query--product pairs, each labeled as \textit{Exact Match}, \textit{Substitutes}, \textit{Complementary}, or \textit{Irrelevant}. This fine-grained annotation facilitates the construction of high-quality positive and negative pairs, which are instrumental during the alignment phase.

\subsection{Behavioral and Implicit Feedback Data}

% \paragraph{Order Data.} 
We first curate a base corpus of search queries and corresponding product results from a continuous timeframe of the platform’s logs. High-confidence query–product relevance labels were derived from aggregated user behavior (e.g., clicks, add-to-cart, purchases) and manual annotations for a subset of query–product pairs. To account for both popular and long-tail queries, we sample queries from a mix of frequency tiers. The resulting dataset contains a diverse range of queries (from single-word to long descriptive phrases) and their relevant products.

\subsection{Recommendation-Based Substitutes}
We leverage co-click signals from the recommendation system to identify product pairs that users frequently consider as interchangeable substitutes. These signals provide a valuable source of domain-specific semantic supervision, especially for capturing substitute patterns that are difficult to infer from textual similarity alone.

We incorporate this data differently across training stages. In the first-stage contrastive training, for branded queries, recommendation-derived substitute products are treated as positives. This encourages the model to place functionally interchangeable products close to the query in the embedding space, thereby improving substitute retrieval. In the subsequent alignment stage, we use a more fine-grained supervision scheme: exact-match products are treated as the strongest positives, while recommendation-derived products are assigned a substitute-level relevance label. This enables the model to distinguish between exact intent and acceptable substitutes, rather than collapsing both into a single relevance class or not retrieving any substitutes at all.

This supervision is particularly important in e-commerce, where brand and product names often have meanings that differ substantially from their general-language usage. For example, terms such as \textit{Munch}, \textit{Perk}, and \textit{Slurrp} correspond to specific food or beverage products in the e-commerce domain, whereas their general-language meanings are unrelated. Incorporating recommendation-based substitutes helps adapt the model's semantic distribution to such domain-specific meanings, enabling it to learn the association of these terms with the corresponding product categories rather than their generic linguistic interpretations.

\subsection{Session-Based Query Reformulation}

User search behavior often involves iterative refinement of queries within a single session~\citep{chen2019investigating,dehghani2017learning}. Initial queries may not yield satisfactory results, prompting users to reformulate their queries to better express their information needs. By analyzing session data, we can capture these reformulations and map the final products that users engage with to all preceding query variations within the session. Session-based query reformulation data in training helps our retrieval model become more adept at understanding user behavior and delivering results that align with the user's evolving search intent.

\subsection{Data Augmentation Strategies}
% \paragraph{\textbf{Synonym Expansion.}} 
\paragraph{\normalfont\textit{Synonym Expansion.}} To address the semantic variability in product ecosystem, we employ LLMs to generate synonyms for product attributes. For instance, the pattern \emph{"checkered"} may be expanded to \emph{"plaid", "gingham", "grid",} and \emph{"checks."} As part of the adversarial perturbation discussed in the next section, these attributes are prompted to these models again to generate search-like queries; and help improve the model's recall in retrieval tasks.

% \paragraph{\textbf{Adversarial Perturbation.}} 
\paragraph{\normalfont\textit{Adversarial Perturbation.}} We further augment the dataset with adversarially perturbed queries (and occasionally documents) to improve the model’s robustness. These include character-level noise (simulated typos, missing or added characters, e.g., \emph{“iphne 13”} for \emph{“iphone 13”}), word-level substitutions with synonyms or related terms (e.g., \emph{“sofa cover”} $\to$ \emph{“couch cover”}), and slight grammar or phrasing changes. Training on such examples (with the same relevance labels as the original queries) reduces the model’s sensitivity to superficial text changes like spelling variations, synonyms, and semantically varying terms. 

% \paragraph{\textbf{Multilingual and Transliteration Data.}} 
\paragraph{\normalfont\textit{Multilingual and Transliteration Data.}} We also generate datasets for multilingual coverage across the Indian subcontinent, including transliterations to English script (e.g., Hindi terms like \emph{``parmal''} for puffed rice, \emph{``kakdi''} for cucumber), global noun understanding, and functional intent variations.

\section{Experimental Setup}

\subsection{Implementation Details}
\label{sec:impl}

\paragraph{Stage 1: Contrastive Learning.} We initialize from Qwen3-Embedding-4B~\citep{qwen2025embedding} with LoRA ($r\!=\!32$, $\alpha\!=\!64$). Training uses a cosine learning rate scheduler with peak rate $2 \times 10^{-5}$ and 10\% warmup. Temperature $\tau = 0.02$, margin $\delta = 0.1$. Stage~1 is trained on approximately 7M query-product pairs.

% Per-device batch size is 1{,}024 across 8 NVIDIA H200 GPUs, yielding a global batch of 8{,}192 with cross-device negative sharing. 
\paragraph{Stage 2: Alignment Learning.} From the Stage~1 checkpoint, we continue LoRA fine-tuning with learning rate $5 \times 10^{-7}$ (constant schedule, 10\% warmup). Hyperparameters: $\beta = 0.1$, $\tau_a = 0.01$. The alignment stage uses a smaller graded-preference dataset of approximately 2M examples.

\paragraph{Distributed Training with Cross-Device Negatives.}

For both stages, we all-gather query and product embeddings across GPUs before the query--product similarity matrix computation, enabling cross-device in-batch negatives. With 8 NVIDIA H200 GPUs and per-device batch size 1{,}024, the effective batch is 8{,}192 query--product pairs. The gather preserves autograd only for the local rank's slice while treating other ranks' embeddings as detached negatives; DDP's all-reduce makes this equivalent to a full-batch gradient update.

\subsection{Baselines}

We compare against the Production Baseline retrieval model, the previously deployed semantic retrieval system in our search pipeline. This baseline is a Siamese BERT-based dual encoder with a shared encoder for query and product representations. The encoder uses 6 transformer layers with a hidden size of 256, and the first \texttt{[CLS]} token representation is used as the embedding. It is trained with a standard InfoNCE contrastive objective using in-batch negatives, where each batch provides $n-1$ negatives per query. The model is trained with large batches, roughly 50K.

\section{Evaluation}
\label{sec:results}

We evaluate on a large-scale benchmark of ${\sim}$25{,}000 queries sampled across frequency tiers (Head, Torso-High, Torso-Low, Tail, Once-Only) and business units, with human-annotated relevance labels. We organize the experimental evaluation into four blocks: (i) an ablation study to quantify the contribution of each modeling component; (ii) overall retrieval across the platform; (iii) a query-segment and business-unit-level breakdown; and (iv) online A/B test metrics.

We report deployment-relevant retrieval metrics that jointly evaluate exact-intent retrieval and graded relevance ordering. MAP@8 measures the extent to which perfect matches appear early in the top-8 results, while AUC measures corpus-level discrimination between perfect matches and non-perfect candidates. For MAP@8 and AUC, we binarize the labels by treating perfect matches as positives and all remaining relevance bands as negatives. NDCG@8 is computed with graded relevance gains over all relevance bands, thereby evaluating whether the ranking appropriately orders different relevance grades. Together, these metrics provide a holistic view of retrieval quality by capturing both perfect-match identification and meaningful relevance gradation. 

\newcommand{\gain}[1]{{\scriptsize\,|\,+#1}}
\begin{table*}[p]
    \centering
    \small

    % ── Tab 1: Ablation (full width) ─────────────────────────────────────────
    \setlength{\tabcolsep}{5pt}
    \begin{threeparttable}
    \begin{tabular}{rlccc}
    \toprule
    \# & \textbf{Configuration} & \textbf{MAP@8} & \textbf{NDCG@8} & \textbf{AUC} \\
    \midrule
    1 & Qwen3-Embedding-0.6B (OOTB) & 0.5454 & 0.7801 & 0.6634 \\
    2 & + Stage~1: contrastive fine-tune (LoRA $r{=}8,\alpha{=}16$, 256D, initial data) & 0.6697 & 0.8571 & 0.7460 \\
    3 & + LoRA capacity (LoRA $r{=}32,\alpha{=}64$) & 0.6740 & 0.8628 & 0.7669 \\
    4 & + Backbone scaling: Qwen3-Embedding-4B & 0.6910 & 0.8732 & 0.7936 \\
    5 & + Contrastive loss with false-negative margin mask & 0.6918 & 0.8756 & 0.7972 \\
    6 & + Stage~2: Contrastive loss with hard-negatives & 0.6968 & 0.8781 & 0.8044 \\
    7 & + ROAR ($g \in [2,7]$) & 0.7083 & 0.8838 & 0.8212 \\
    8 & + Recommendation-based and session-reformulation data & 0.7131 & 0.8920 & 0.8350 \\
    9 & + Data augmentation strategies (\textbf{\sysname @256D}) & \textbf{0.7163} & \textbf{0.8974} & \textbf{0.8441} \\
    \midrule
    \multicolumn{2}{l}{\emph{Variation:} \sysname @2560D} & 0.7213 & 0.8999 & 0.8430 \\
    \multicolumn{2}{l}{\textbf{Production Baseline}} & 0.6606 & 0.8618 & 0.7128 \\
    \bottomrule
    \end{tabular}
    \par%\vspace{1pt}
    \caption{\textmd{Incremental ablation of the proposed retrieval system. Each row adds one modeling, training, or data component to the previous configuration, showing cumulative effects on offline retrieval quality.}}
    \label{tab:hl_ablation}
    \end{threeparttable}

    \vspace{2\baselineskip}\vfill

    % ── Tab 2: Headline results (full width) ─────────────────────────────────
    \setlength{\tabcolsep}{6pt}
    \begin{threeparttable}
    \begin{tabular}{lccc}
    \toprule
    \textbf{Model} & \textbf{MAP@8} & \textbf{NDCG@8} & \textbf{AUC} \\
    \midrule
    Production Baseline & 0.6606 & 0.8618 & 0.7128 \\
    \sysname @256D      & 0.7163\gain{5.57} & 0.8974\gain{3.56} & \textbf{0.8441\gain{13.13}} \\
    \sysname @2560D     & \textbf{0.7213\gain{6.07}} & \textbf{0.8999\gain{3.81}} & 0.8430\gain{13.02} \\
    \bottomrule
    \end{tabular}
    \par%\vspace{1pt}
    \caption{\textmd{Headline offline retrieval results comparing the production baseline with the final \sysname{} variants.}}
    \label{tab:main_hl}
    \end{threeparttable}
    \vspace{2\baselineskip}\vfill

    % ── Tab 3: Segment breakdown (full width) ────────────────────────────────
    \setlength{\tabcolsep}{7pt}
    \begin{threeparttable}
    \begin{tabular}{lcccc}
    \toprule
     & \multicolumn{2}{c}{\textbf{MAP@8}} & \multicolumn{2}{c}{\textbf{NDCG@8}} \\
    \cmidrule(lr){2-3} \cmidrule(lr){4-5}
    \textbf{Segment} & \textbf{Baseline} & \textbf{\sysname} & \textbf{Baseline} & \textbf{\sysname} \\
    \midrule
    HEAD        & 0.9412 & 0.9420\textbf{\gain{0.08}} & 0.9618 & 0.9637\textbf{\gain{0.18}} \\
    TORSO HIGH  & 0.9435 & 0.9628\textbf{\gain{1.92}} & 0.9788 & 0.9867\textbf{\gain{0.79}} \\
    TORSO LOW   & 0.8552 & 0.8913\textbf{\gain{3.60}} & 0.9517 & 0.9609\textbf{\gain{0.92}} \\
    TAIL        & 0.7767 & 0.8113\textbf{\gain{3.46}} & 0.9279 & 0.9428\textbf{\gain{1.48}} \\
    ONCE ONLY   & 0.4403 & 0.5257\textbf{\gain{8.54}} & 0.7513 & 0.8192\textbf{\gain{6.79}} \\
    \bottomrule
    \end{tabular}
    \par%\vspace{1pt}
    \caption{\textmd{Retrieval quality by query frequency. \sysname{} preserves head-query performance while improving sparse and once-only queries.}}
    \label{tab:hl_segment}
    \end{threeparttable}
    
    \vspace{2\baselineskip}\vfill

    % ── Tab 4: Business unit (full width) ────────────────────────────────────
    \setlength{\tabcolsep}{7pt}
    \begin{threeparttable}
    \begin{tabular}{lcccc}
    \toprule
     & \multicolumn{2}{c}{\textbf{MAP@8}} & \multicolumn{2}{c}{\textbf{NDCG@8}} \\
    \cmidrule(lr){2-3} \cmidrule(lr){4-5}
    \textbf{Business Unit} & \textbf{Baseline} & \textbf{\sysname} & \textbf{Baseline} & \textbf{\sysname}  \\
    \midrule
    BGM         & 0.6710 & 0.7194\textbf{\gain{4.84}}  & 0.8863 & 0.9103\textbf{\gain{2.41}}  \\
    Electronics & 0.7111 & 0.7630\textbf{\gain{5.19}}  & 0.8962 & 0.9250\textbf{\gain{2.88}}  \\
    Food        & 0.7618 & 0.7940\textbf{\gain{3.22}}  & 0.9374 & 0.9466\textbf{\gain{0.91}}  \\
    Home        & 0.6603 & 0.7282\textbf{\gain{6.79}}  & 0.8539 & 0.8874\textbf{\gain{3.34}}  \\
    Large       & 0.4934 & 0.5330\textbf{\gain{3.96}}  & 0.6504 & 0.6944\textbf{\gain{4.40}}  \\
    Lifestyle   & 0.5692 & 0.6742\textbf{\gain{10.50}} & 0.8068 & 0.8850\textbf{\gain{7.82}}  \\
    Mobiles     & 0.4301 & 0.5742\textbf{\gain{14.41}} & 0.7391 & 0.9109\textbf{\gain{17.18}} \\
    Ambiguous   & 0.4839 & 0.5486\textbf{\gain{6.46}}  & 0.6953 & 0.7563\textbf{\gain{6.10}}  \\
    \bottomrule
    \end{tabular}
    \par%\vspace{1pt}
    \caption{\textmd{Retrieval quality across business units. \sysname{} improves both MAP@8 and NDCG@8 consistently across diverse catalog segments.}}
    \label{tab:hl_bu}
    \end{threeparttable}
    
    \vspace{2\baselineskip}\vfill

    % ── Tab 5 (left) │ Tab 6 (right): A/B results and serving performance ────
    \begin{minipage}[t]{0.47\textwidth}
        \centering
        \setlength{\tabcolsep}{10pt}
        \begin{tabular}{lc}
            \toprule
            \textbf{Online Metric} & \textbf{Lift over Baseline} \\
            \midrule
            Click-through Rate (CTR) & +2.39\% \\
            Add-to-Cart (ATC)        & +4.58\% \\
            Orders                   & +2.62\% \\
            \bottomrule
        \end{tabular}
        % \par\smallskip
        \caption{\textmd{Online A/B test impact on business metrics.}}
        \label{tab:online_ab_results}
    \end{minipage}
    \hfill
    \begin{minipage}[t]{0.47\textwidth}
        \centering
        \setlength{\tabcolsep}{8pt}
        \begin{tabular}{lccc}
            \toprule
            \textbf{Serving Load} & \textbf{QPS} & \textbf{P50} & \textbf{P99} \\
            \midrule
            Nominal load     & 200 & 40 ms & 135 ms \\
            Peak tested load & 680 & 80 ms & 236 ms \\
            % ~ & ~ & ~ & ~ \\
            \bottomrule
        \end{tabular}
        % \par\smallskip
        \caption{\textmd{Serving performance of Qwen3-4B with vLLM on a \texttt{1g.18GB} H200 MIG partition.}}
        \label{tab:nfr_metrics}
    \end{minipage}

    \vfill
    \vspace{5mm}

    {\footnotesize
    Unless stated otherwise, entries of the form $x~{\scriptsize |~+y}$ report the absolute metric score $x$ with the absolute percentage-point gain $y$ over the corresponding Production Baseline. Tables without inline gains report absolute metric scores only.
    }
    
\end{table*}

\section{Results and Analysis}

All reported gains are absolute metric-point differences over the corresponding baseline, not relative percentage improvements.

\subsection{Consolidated Ablation Study}
\label{sec:hl_ablation}

Table~\ref{tab:hl_ablation} reports an incremental ablation, starting from an out-of-the-box Qwen3-0.6B model and culminating in the final deployed \sysname. Each row adds exactly one factor to the previous configuration. We highlight three observations: (i) Stage~1 contrastive fine-tuning on human-annotated and behavioral data provides the bulk of the improvement over OOTB; (ii) scaling the backbone from 0.6B to 4B yields additional gains (+1.70 MAP@8 points, +2.67 AUC points); and (iii) Stage~2 ROAR over contrastive loss with hard negatives gives the largest single algorithmic gain (+1.15 / +0.57 / +1.68 points on MAP@8 / NDCG@8 / AUC, respectively) at no indexing cost. Behavioral and augmented training data (rows 8--9) provide further gains, particularly on tail queries.

\subsection{Platform-level Retrieval Results}
\label{sec:results_hl_main}

Table~\ref{tab:main_hl} reports headline retrieval quality on our evaluation set. \sysname, deployed at 256 dimensions through MRL, improves MAP@8 by +5.57 points, NDCG@8 by +3.56 points, and AUC by +13.13 points over the Production Baseline. The 2560D variant achieves marginally higher MAP@8 (+0.50 points over 256D) but slightly lower AUC, motivating the 256D deployment choice.

\subsection{Performance by Query Segment (HTTTO)}

Table~\ref{tab:hl_segment} breaks down retrieval quality by query-frequency. Gains increase with query rarity: head queries gain only +0.08 MAP@8 points, reflecting saturated performance, while once-only queries gain +8.54 points. This supports our hypothesis that larger models, data augmentation, multi-pool contrastive learning, and graded alignment mainly benefit under-represented query distributions.

\subsection{Performance by Business Unit}

Table~\ref{tab:hl_bu} reports business-unit-level gains across eight major verticals. Mobiles (+14.41 MAP@8 points) and Lifestyle (+10.50 points) improve the most, reflecting the value of better brand, specification, and style-attribute understanding. Smaller-catalog verticals such as Food (+3.22 points) show smaller gains, consistent with more saturated text-based discoverability.

\subsection{Online A/B Test Results}

Table~\ref{tab:online_ab_results} reports the online impact of deploying \sysname. The system improves CTR by +2.39 points, ATC by +4.58 points, and Orders by +2.62 points over the production baseline.

\section{Inference and Non-functional Metrics}

We deployed the Qwen3-4B model on NVIDIA H200 GPU using a \texttt{1g.18GB} MIG partition. The model has a GPU memory footprint of approximately 7.5GB in \texttt{bfloat16}, enabling deployment within a single MIG slice. Inference is served using vLLM~\citep{kwon2023efficient}. Table~\ref{tab:nfr_metrics} reports the latency-throughput trade-off.

\section{Conclusion}

In this work, we present an LLM-based semantic matching system for search at Flipkart. We describe the key components required to build and deploy this system, including the training data construction pipeline, model scaling strategy, two-stage optimization recipe, alignment with graded relevance signals, and serving for low-latency retrieval. Through extensive offline experiments, we show that each modeling and data intervention contributes to consistent improvements over the production baseline across retrieval metrics, query-frequency segments, and business units. Finally, our online A/B test demonstrates that these offline gains translate into measurable business impact, improving CTR by 2.39 points, ATC by 4.58 points, and Orders by 2.62 points over the baseline. These results establish that large embedding models, when carefully adapted and optimized for domain-specific retrieval, can substantially improve search relevance at production scale.

\section*{Limitations}
\label{sec:limitations}

This work is evaluated on proprietary data from a single large-scale e-commerce platform, which limits direct comparison with public benchmarks and leaves cross-domain generalization to future work. Our offline evaluation relies on relevance judgments from an internal annotation pipeline; since the annotation guidelines, label distribution, and inter-annotator agreement are not publicly released, the results are not fully reproducible outside our setting. The evaluation data is also primarily English, with some transliterated Indian-language queries, and does not separately measure performance on fully Indian-language queries. Finally, the online A/B test covers a three-week deployment window, so longer-term effects on user behavior, catalog freshness, and selection diversity remain unstudied. The proposed alignment stage also assumes access to graded group-preference annotations, which may require additional labeling or synthetic data in annotation-scarce settings.

\section*{Ethical Considerations}
\label{sec:ethics}
The training corpus is derived from logged user search behaviour on our platform. All logs are aggregated and anonymised at ingestion, and contain no personally identifiable information. Annotation is performed by an internal labelling team under standard data-handling agreements. The deployed model serves only retrieval-stage candidates and does not make purchase decisions or interact directly with end users beyond surfacing search results. We release no proprietary data, model weights, or internal benchmarks. Standard search-relevance failures (e.g., mistakes on rare brands, regional terms, or long-tail intents) are surfaced through monitoring and addressed through the data-augmentation pipeline.

\bibliography{main}

@article{li2021learning,
  title={From Semantic Retrieval to Pairwise Ranking: Applying Deep Learning in E-commerce Search},
  author={Li, Rui and Jiang, Yunjiang and Yang, Wenyun and Tang, Guoyu and Wang, Songlin and Ma, Chaoyi and He, Wei and Xiong, Xi and Xiao, Yun and Zhao, Eric Yihong},
  journal={arXiv preprint arXiv:2103.12982},
  year={2021}
}

@inproceedings{nogueira2019passage,
  title={Passage re-ranking with BERT},
  author={Nogueira, Rodrigo and Cho, Kyunghyun},
  booktitle={Proceedings of the 2019 Conference of the North American Chapter of the Association for Computational Linguistics: Human Language Technologies},
  year={2019}
}

@inproceedings{karpukhin2020dense,
  title={Dense passage retrieval for open-domain question answering},
  author={Karpukhin, Vladimir and Oguz, Barlas and Min, Sewon and Lewis, Patrick and Wu, Ledell and Edunov, Sergey and Chen, Danqi and Yih, Wen-tau},
  booktitle={Proceedings of the 2020 Conference on Empirical Methods in Natural Language Processing (EMNLP)},
  pages={6769--6781},
  year={2020}
}

@article{hong2024orpo,
  title={ORPO: Monolithic Preference Optimization without Reference Model},
  author={Hong, Jiwoo and Lee, Noah and Thorne, James},
  journal={arXiv preprint arXiv:2403.07691},
  year={2024}
}

@inproceedings{xiong2020approximate,
  title     = {Approximate Nearest Neighbor Negative Contrastive Learning for Dense Text Retrieval},
  author    = {Xiong, Lee and Xiong, Chenyan and Li, Ye and Tang, Kwok-Fung and Liu, Jialin and Bennett, Paul N. and Ahmed, Junaid and Overwijk, Arnold},
  booktitle = {International Conference on Learning Representations},
  year      = {2021},
  url       = {https://arxiv.org/abs/2007.00808}
}

@inproceedings{yao2021learning,
  title={Learning a Product Relevance Model from Click-Through Data in E-Commerce},
  author={Yao, Shaowei and Tan, Jiwei and Chen, Xi and Yang, Keping and Xiao, Rong and Deng, Hongbo and Wan, Xiaojun},
  booktitle={Proceedings of the Web Conference 2021},
  pages={2890--2899},
  year={2021},
  organization={ACM},
  doi={10.1145/3442381.3450078}
}

@inproceedings{chen2019investigating,
  title={Investigating Query Reformulation Behavior of Search Users},
  author={Chen, Jia and others},
  booktitle={Chinese Conference on Information Retrieval},
  pages={34--46},
  year={2019},
  organization={Springer}
}

@inproceedings{dehghani2017learning,
  title={Learning to Attend, Copy, and Generate for Session-Based Query Suggestion},
  author={Dehghani, Mostafa and Rothe, Sascha and Alfonseca, Enrique and Fleury, Pascal},
  booktitle={Proceedings of the 2017 ACM on Conference on Information and Knowledge Management},
  pages={1747--1756},
  year={2017},
  organization={ACM}
}

@inproceedings{joachims2017unbiased,
  title={Unbiased Learning-to-Rank with Biased Feedback},
  author={Joachims, Thorsten and Swaminathan, Adith and Schnabel, Tobias},
  booktitle={Proceedings of the Tenth ACM International Conference on Web Search and Data Mining},
  pages={781--789},
  year={2017},
  organization={ACM}
}

@inproceedings{kusupati2022matryoshka,
  title={Matryoshka Representation Learning},
  author={Kusupati, Aditya and Bhatt, Gantavya and Rber, Aniket and Wallingford, Matthew and Sinha, Aditya and Patil, Vivek and Simhadri, Harsha Vardhan and Jain, Prateek},
  booktitle={Advances in Neural Information Processing Systems},
  volume={35},
  pages={30233--30249},
  year={2022}
}

@inproceedings{hu2022lora,
  title={{LoRA}: Low-Rank Adaptation of Large Language Models},
  author={Hu, Edward J. and Shen, Yelong and Wallis, Phillip and Allen-Zhu, Zeyuan and Li, Yuanzhi and Wang, Shean and Wang, Lu and Chen, Weizhu},
  booktitle={International Conference on Learning Representations},
  year={2022},
  url={https://arxiv.org/abs/2106.09685}
}

@misc{qwen2025embedding,
  title={Qwen3 Embedding: Advancing Text Embedding and Reranking Through LLMs},
  author={Qwen Team},
  year={2025},
  howpublished={\url{https://qwenlm.github.io/blog/qwen3-embedding/}},
  note={Accessed: 2025-06-01}
}

@inproceedings{reimers2019sentence,
  title={Sentence-{BERT}: Sentence Embeddings using Siamese {BERT}-Networks},
  author={Reimers, Nils and Gurevych, Iryna},
  booktitle={Proceedings of the 2019 Conference on Empirical Methods in Natural Language Processing (EMNLP)},
  pages={3982--3992},
  year={2019}
}

@inproceedings{khattab2020colbert,
  title={{ColBERT}: Efficient and Effective Passage Search via Contextualized Late Interaction over {BERT}},
  author={Khattab, Omar and Zaharia, Matei},
  booktitle={Proceedings of the 43rd International ACM SIGIR Conference on Research and Development in Information Retrieval},
  pages={39--48},
  year={2020}
}

@article{ni2022large,
  title={Large Dual Encoders Are Generalizable Retrievers},
  author={Ni, Jianmo and Qu, Chen and Lu, Jing and Dai, Zhuyun and Hern{\'a}ndez, Gustavo and Jindal, Madhav and Feng, Yun and Gopi, Shankar and Cer, Daniel and others},
  journal={arXiv preprint arXiv:2112.07899},
  year={2022}
}

@article{wang2024improving,
  title={Improving Text Embeddings with Large Language Models},
  author={Wang, Liang and Yang, Nan and Huang, Xiaolong and Yang, Linjun and Majumder, Rangan and Wei, Furu},
  journal={arXiv preprint arXiv:2401.00368},
  year={2024}
}

@article{lee2024nv,
  title={{NV-Embed}: Improved Techniques for Training {LLM}s as Generalist Embedding Models},
  author={Lee, Chankyu and Roy, Rajarshi and Xu, Menber and Raiman, Jonathan and Shoeybi, Mohammad and Catanzaro, Bryan and Wei, Wei},
  journal={arXiv preprint arXiv:2405.17428},
  year={2024}
}

@article{muennighoff2024generative,
  title={Generative Representational Instruction Tuning},
  author={Muennighoff, Niklas and Su, Hongjin and Wang, Liang and Yang, Nan and Wei, Furu and Yu, Tao and Singh, Amanpreet and Kiela, Douwe},
  journal={arXiv preprint arXiv:2402.09906},
  year={2024}
}

@inproceedings{huang2020embedding,
  title={Embedding-based Retrieval in Facebook Search},
  author={Huang, Jui-Ting and Sharma, Ashish and Sun, Shuying and Xia, Li and Zhang, David and Pronin, Philip and Padmanabhan, Janani and Ottaviano, Giuseppe and Yang, Linjun},
  booktitle={Proceedings of the 26th ACM SIGKDD International Conference on Knowledge Discovery and Data Mining},
  pages={2553--2561},
  year={2020}
}

@inproceedings{nigam2019semantic,
  title={Semantic Product Search},
  author={Nigam, Priyanka and Song, Yiwei and Mohan, Vijai and Lakshman, Viren and Ding, Weitian and Shinber, Ankit and Gagber, Rahul and Bhatia, Saurabh},
  booktitle={Proceedings of the 25th ACM SIGKDD International Conference on Knowledge Discovery and Data Mining},
  pages={2876--2885},
  year={2019}
}

@inproceedings{fan2019mobius,
  title={{MOBIUS}: Towards the Next Generation of Query-Ad Matching in Baidu's Sponsored Search},
  author={Fan, Miao and Guo, Jiacheng and Zhu, Shuai and Miao, Shuo and Sun, Mingming and Li, Ping},
  booktitle={Proceedings of the 25th ACM SIGKDD International Conference on Knowledge Discovery and Data Mining},
  pages={2509--2517},
  year={2019}
}

@article{qu2021rocketqa,
  title={{RocketQA}: An Optimized Training Approach to Dense Passage Retrieval for Open-Domain Question Answering},
  author={Qu, Yingqi and Ding, Yuchen and Liu, Jing and Liu, Kai and Ren, Ruiyang and Zhao, Wayne Xin and Dong, Daxiang and Wu, Hua and Wang, Haifeng},
  journal={Proceedings of the 2021 Conference of the North American Chapter of the Association for Computational Linguistics},
  pages={5835--5847},
  year={2021}
}

@inproceedings{reddy2022shopping,
  title={Shopping Queries Dataset: A Large-Scale {ESCI} Benchmark for Improving Product Search},
  author={Reddy, Chandan K. and Magnani, Llu{\'i}s and Feng, Yan and Guo, Liyun and Ren, Peng},
  booktitle={Proceedings of the 28th ACM SIGKDD International Conference on Knowledge Discovery and Data Mining},
  pages={3553--3562},
  year={2022}
}

@inproceedings{chen2022wands,
  title={{WANDS}: Dataset for Product Search Relevance Assessment},
  author={Chen, Yan and Liu, Shujian and He, Zheng and Subudhi, Aman and Tou, Yingyu Rachel},
  booktitle={Proceedings of the 45th European Conference on Information Retrieval},
  year={2022}
}

@inproceedings{bonifacio2022inpars,
  title={{InPars}: Data Augmentation for Information Retrieval using Large Language Models},
  author={Bonifacio, Luiz and Abonizio, Hugo and Fadaee, Marzieh and Nogueira, Rodrigo},
  booktitle={Proceedings of the 45th International ACM SIGIR Conference on Research and Development in Information Retrieval},
  pages={2316--2320},
  year={2022}
}

@article{dai2023promptagator,
  title={Promptagator: Few-shot Dense Retrieval From 8 Examples},
  author={Dai, Zhuyun and Zhao, Vincent Y. and Ma, Ji and Luan, Yi and Ni, Jianmo and Lu, Jing and Baber, Anton and Callan, Jamie and Cer, Daniel},
  journal={Proceedings of the 11th International Conference on Learning Representations},
  year={2023}
}

@inproceedings{kwon2023efficient,
  title     = {Efficient Memory Management for Large Language Model Serving with {PagedAttention}},
  author    = {Kwon, Woosuk and Li, Zhuohan and Zhuang, Siyuan and Sheng, Ying and Zheng, Lianmin and Yu, Cody Hao and Gonzalez, Joseph E. and Zhang, Hao and Stoica, Ion},
  booktitle = {Proceedings of the 29th Symposium on Operating Systems Principles ({SOSP})},
  year      = {2023}
}

@article{bradley1952rank,
  title={Rank analysis of incomplete block designs: {I}. {T}he method of paired comparisons},
  author={Bradley, Ralph Allan and Terry, Milton E},
  journal={Biometrika},
  volume={39},
  number={3/4},
  pages={324--345},
  year={1952},
  publisher={JSTOR}
}
\end{document}